\documentstyle[11pt,aaspp4]{article}

\newcommand\MSUNYR{\rm M_{\odot}\,yr^{-1}}
\newcommand\MSUN{\rm M_{\odot}}
\newcommand\LSUN{\rm L_{\odot}}
\newcommand\RSUN{\rm R_{\odot}}
\newcommand\Mdot{ \dot{M}}
\newcommand\etal{et al. }
\newcommand\be {\begin{equation}}
\newcommand\en{\end{equation}}
\newcommand\cm{\rm cm}

\newcommand\AU{\rm AU}

\def\'#1{\ifx#1i{\accent"13\i}\else{\accent"13#1}\fi}

\def\10alamenos#1{10$^{-#1}$}
\def\10ala#1{10$^{#1}$}

\def\prom#1{\langle #1\rangle}

\def\alb{\omega_\nu}

\def\cm2g{\rm cm^2 \ g^{-1}}

\begin{document}

\title
{ACCRETION DISKS AROUND YOUNG OBJECTS. III. GRAIN GROWTH}

\author{Paola D'Alessio \altaffilmark{1,2}, Nuria Calvet \altaffilmark{3}, and Lee Hartmann 
\altaffilmark{3}}
\altaffiltext{1}{Department of Astrophysics,
American Museum of Natural History, Central Park West at 79th Street, New York,
NY10024-5192; Electronic mail: paola@amnh.org}
\altaffiltext{2}{Instituto de Astronom\'{\i}a,
UNAM, Ap. Postal 70-264, Cd. Universitaria, 04510 M\'exico D.F.,
M\'exico}
\altaffiltext{3}{Harvard-Smithsonian Center for Astrophysics, 60 Garden St.,
Cambridge, MA 02138, USA;
Electronic mail: hartmann@cfa.harvard.edu, ncalvet@cfa.harvard.edu}

\begin{abstract}
We present detailed models of irradiated T Tauri disks including 
dust grain growth
with power-law size distributions.  The models assume complete mixing between
dust and gas and solve for the vertical disk structure self-consistently
including the heating effects of stellar irradiation as well as local
viscous heating.  For a given total dust mass, grain growth is found 
to decrease the vertical height of the surface where the optical 
depth to the stellar radiation becomes unit 
and thus
the local irradiation heating, while increasing the disk emission at mm
and sub-mm wavelengths.  The resulting disk models 
are less geometrically thick
than our previous models assuming interstellar medium dust, and agree
better with observed spectral energy distributions and images of edge-on 
disks, like HK Tau/c and HH 30.
The implications of models with grain growth for determining disk masses
from long-wavelength emission are considered.
\end{abstract}

\keywords{Physical data and processes: accretion, accretion disks ---  stars:
circumstellar matter,  formation, pre-main sequence}

\section{Introduction}
\label{sec_intro}

It has been clear for more than a decade that the absorbing surfaces or
photospheres of T Tauri disks had to be ``flared'',
i.e. curved upward away from the disk midplane, to explain the typical observed
spectral energy distribution at infrared wavelengths (\cite{KH87}).
Because of flaring, the irradiation of the disk by the central star tends to
dominate the disk heating at large radii (\cite{CMPD}).  Since the amount
of flaring depends upon the temperature structure, which in turn depends upon
the flaring-dependent irradiation, the self-consistent structure in general
must be solved iteratively (\cite{KH87}; \cite{D96};
D'Alessio et al. 1998, 1999a).  The dust grains control the absorption and 
emission
of radiation and therefore 
the knowledge of 
their properties help determine the disk structure.

In a previous paper (D'Alessio et al. 1999b $=$ Paper II), we constructed 
models for T Tauri disks in which the vertical disk structure was 
self-consistently
solved assuming well-mixed dust with properties characteristic of 
interstellar grains.
We found that the resulting models were too geometrically thick, which 
is also the case of the two-layer flared 
disk models of  Chiang \& Goldreich (1997, 1999). Our models 
also exhibited too
little mm- and sub-mm wavelength emission, in comparison with observations.
These deficiencies can in principle be relieved or eliminated by models 
including
grain growth and settling to the disk midplane, processes which are 
thought to be 
essential precursors to planet formation (e.g., \cite{BHN00}).
Even without settling, grain growth
alone can help improve the comparison with observations, since for a given dust
mass growth can increase the long-wavelength opacity at the same time it 
decreases
the optical-near infrared opacity; the latter effect reduces the apparent 
disk thickness
at short wavelengths, and can even reduce the true disk vertical scale height 
by reducing irradiation heating.  

It is generally thought that some grain growth must have occurred in T Tauri 
disks 
to explain the spectral indices of the long-wavelength disk emission in some 
systems 
(Beckwith \& Sargent 1991 $=$ BS91; \cite{ME94}; \cite{P94} $=$ P94).
Unfortunately, the details of this process are poorly understood.  We are 
therefore
motivated to adopt a parameterized approach with few parameters to make an 
initial
exploration of the effects of dust growth on T Tauri disk structure.

In this paper we present sequences of detailed, physically
self-consistent disk models with power-law distributions of dust particle 
sizes.
We find that modest grain growth substantially improves comparison with 
spectral
energy distributions (SEDs) and optical-near infrared imaging.
Using both observed mm-wave fluxes and spectral slopes as constraints,
we suggest limits on the maximum grain size in some systems.  Our results 
emphasize the
importance of optical depth effects in constraining disk properties, and 
suggest
observational tests which might ultimately lead to more accurate disk masses.
The presence of silicate emission may require differential distributions of 
small
and large dust, i.e., settling of large grains to midplane with small grains
(size $< 10 \ \mu$m) at larger heights. This issue will be addressed in a 
forthcoming paper of this series.

\section{Dust opacities}
\label{sec_opa}

Dust is the most important opacity source in T Tauri disk models, and controls
the temperature distribution within the disk as well as the emissivity at 
long wavelengths.
We adopt the model of dust for disks proposed by P94, with a few variations.
We assume that the optical properties of olivine and orthopyroxene are 
represented
by the ``astronomical silicates'' (Draine \& Lee 1984),i.e., crystalline 
olivine  with dielectric functions given by
\cite{LD93} and \cite{WD00}. However, following P94, the imaginary refractive 
index is taken
to be constant for $\lambda > 800 \ \mu$m, for consistency with laboratory 
measurements
(\cite{CU69}; \cite{HS96}).  The optical constants of water ice are taken 
from \cite{Warren84}.
Refractive indices for troilite are taken from \cite{beg94} 
(see \cite{hen99}) for
$10 < \lambda < 500 \ \mu$m and  from P94  for other wavelengths.  
Finally,
the refractive indices of organics are taken from P94.  The 
mass fractional abundances
of these dust species respect to gas are $\zeta_{sil}=0.0034$,
$\zeta_{tro}=0.000768$, $\zeta_{ice}=0.0056$, and $\zeta_{org}=0.0041$, and
their bulk densities are $\rho_{sil}=3.3$ , $\rho_{tro}=4.83$,
$\rho_{ice}=0.92$ and $\rho_{org}=1.5$ g cm$^{-3}$, respectively, consistent 
with P94.
The sublimation temperature of each of these species is taken from P94.

The efficiency factor $Q_{abs}$ is calculated using a Mie scattering
code (\cite{wis79}, updated in 1996),
assuming that the grains are spheres.
Coagulation tends to produce grains which are not compact spheres, but rather
aggregates with a porous structure and a bulk density that decreases with 
size (\cite{MD88}).
The shapes of these grains are affected by different kind of processes
which may take place in the disk, such as turbulence, settling, radial 
drift, etc.; 
in turn, the optical properties of these aggregates
depend on their complex shapes (e.g., \cite{HS96}; \cite{BHN00}).
However,  in the relatively dense environment of a
circumstellar disk, there are frequent collisions between
particles and a strong interaction  with the surrounding gas, 
tending to produce erosion
and compaction of grains (Miyake \& Nakagawa 1993 $=$ MN93).
In any event, for simplicity and lacking any detailed description of the
shapes and optical properties of grains in turbulent disks,
we assume that the disk grains interact with light as compact 
segregated 
spheres.

The size distribution of dust grains is taken to be a power law 
$n(a) = n_{0} \  a^{-p}$,
where $a$ is the grain radius, $n_0$ is a normalization constant and 
the exponent $p$ is a free parameter.
Approximate justification for this is taken from the calculations of 
Weidenschilling
(1997), who finds a wide range of dust sizes at any given time due to a 
combination
of grain growth and shattering.  We assume a minimum grain radius 
$a_{min}=0.005 \ \mu$m and
different maximum size  $a_{max}$ and exponent $p$.
Table 1 summarizes representative values 
of the opacity and albedo at different wavelengths for various 
$a_{max}$ and $p$.
In this paper we use the following notation: $\kappa_\nu$ is the mass
absorption coefficient (i.e., corresponding to  ``true absorptions''), 
$\chi_\nu$ is the total extinction coefficient 
(i.e., absorptions plus scattering) 
and $w_\nu$ is the albedo (given by $w_\nu=1-\kappa_\nu/\chi_\nu$).
For comparison, we frequently refer to ``ISM-dust'', which corresponds to 
abundances and optical properties from \cite{DL84}, i.e.,  
``astronomical silicates'' 
and graphite, with $a_{max}=$0.25  $\mu$m and $p=$3.5.

\begin{figure}
\plotone{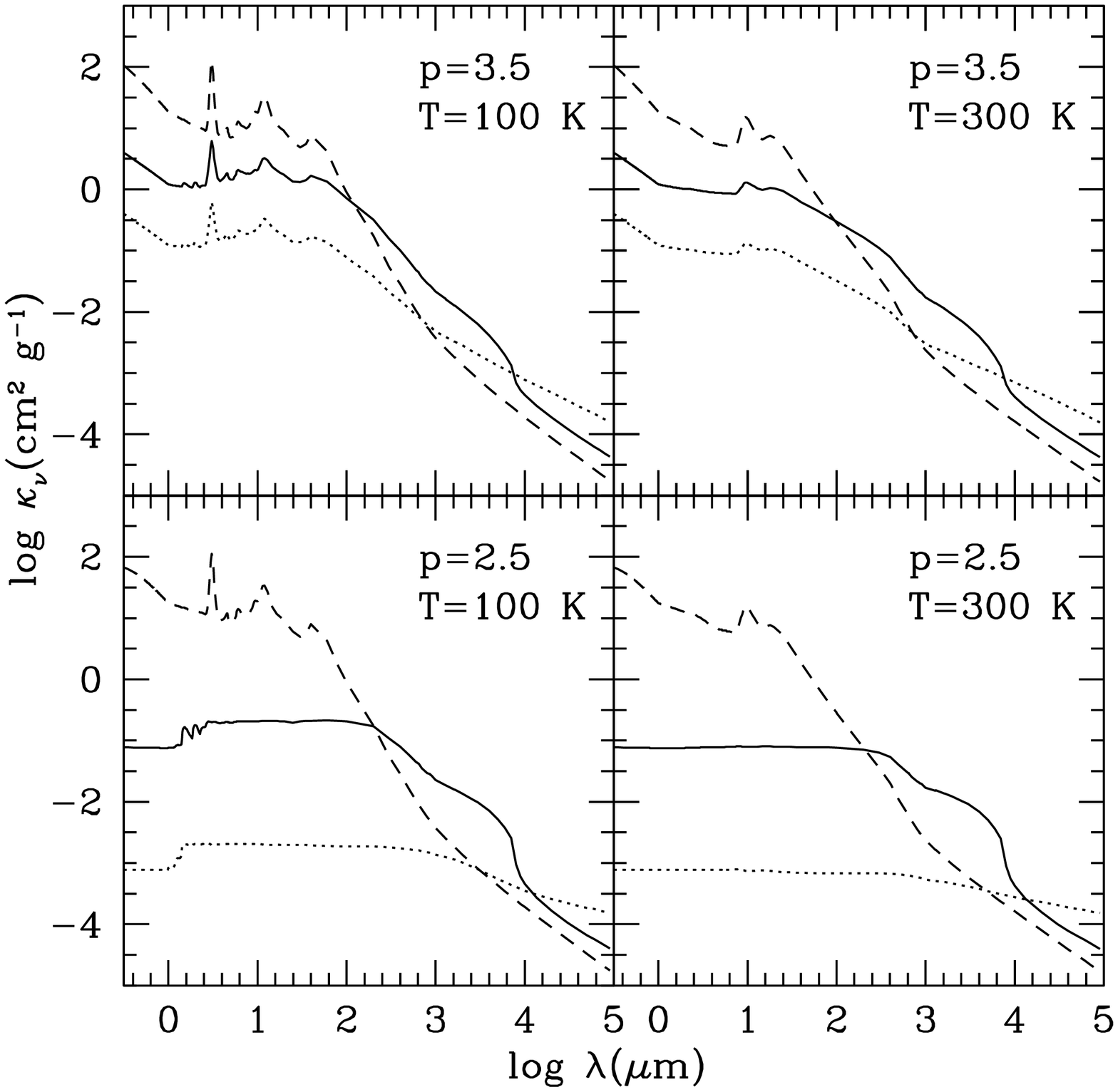}
\caption{Mass absorption coefficient $\kappa_\nu$ vs wavelength
for a grain size distribution characterized by  $p=3.5$ (upper panels) 
and $2.5$ (lower panels).
Each curve corresponds to a different grain maximum size: 
$a_{max}={\rm 1} \ \mu m$ (dashed line), 1 mm (solid line), 
10 cm (dotted line).
The absorption coefficient is shown for $T=$ 100 K 
(i.e, all the different kind of grains are present) and $T=$ 300 K 
(i.e., water ice has been sublimated). 
Each panel is labeled 
by $p$ and $T$.} 
\label{fig_opa_varias}
\end{figure}

Figure \ref{fig_opa_varias} shows the opacity as a function of 
wavelength for dust grain 
size distributions 
with $p=$2.5 and 3.5, and maximum size $a_{max}=$10  $\mu$m, 
1 mm and 10 cm, for temperatures $T=$ 100 and 300 K (i.e., with and 
without water ice, respectively). 
For a given value of $p$, when the maximum grain size increases, 
the opacity at small 
wavelengths ($ \lambda \lesssim a_{max}$) decreases and the opacity 
at larger wavelengths increases.
Since the dust to gas mass ratio is assumed to be constant and the mass in
grains is dominated by the bigger grains if $p <$ 4 (MN93),
a larger $a_{max}$ implies a smaller mass in small grains which are the most 
efficient absorbing radiation in the visible and near-IR.
At the same time, as $ a_{max}$ increases, the transition between
the geometrical and optical regimes occurs at longer
wavelengths ($\sim 2 \pi a_{max}$), so that the
opacity at $\lambda \sim a_{max}$ increases.
In Table 1 we show the Planck mean opacities 
characteristic of the 
stellar radiation $\kappa_P^*$ 
and $\chi_P^*$, 
both calculated using the Planck function evaluated at a typical stellar 
temperature $T_*=4000$ K 
(see paper II). These values are similar to the monochromatic coefficient 
evaluated at 
$\lambda \sim 1 \ \mu m$, and are used to calculate the disk heating by 
stellar radiation 
(see paper II for details).  The table illustrates how these mean 
opacities also 
decrease with $a_{max}$. The consequences of this 
behavior for the disk structure 
and emission properties will be discussed in \S \ref{sec_results}.

\begin{figure}
\plotone{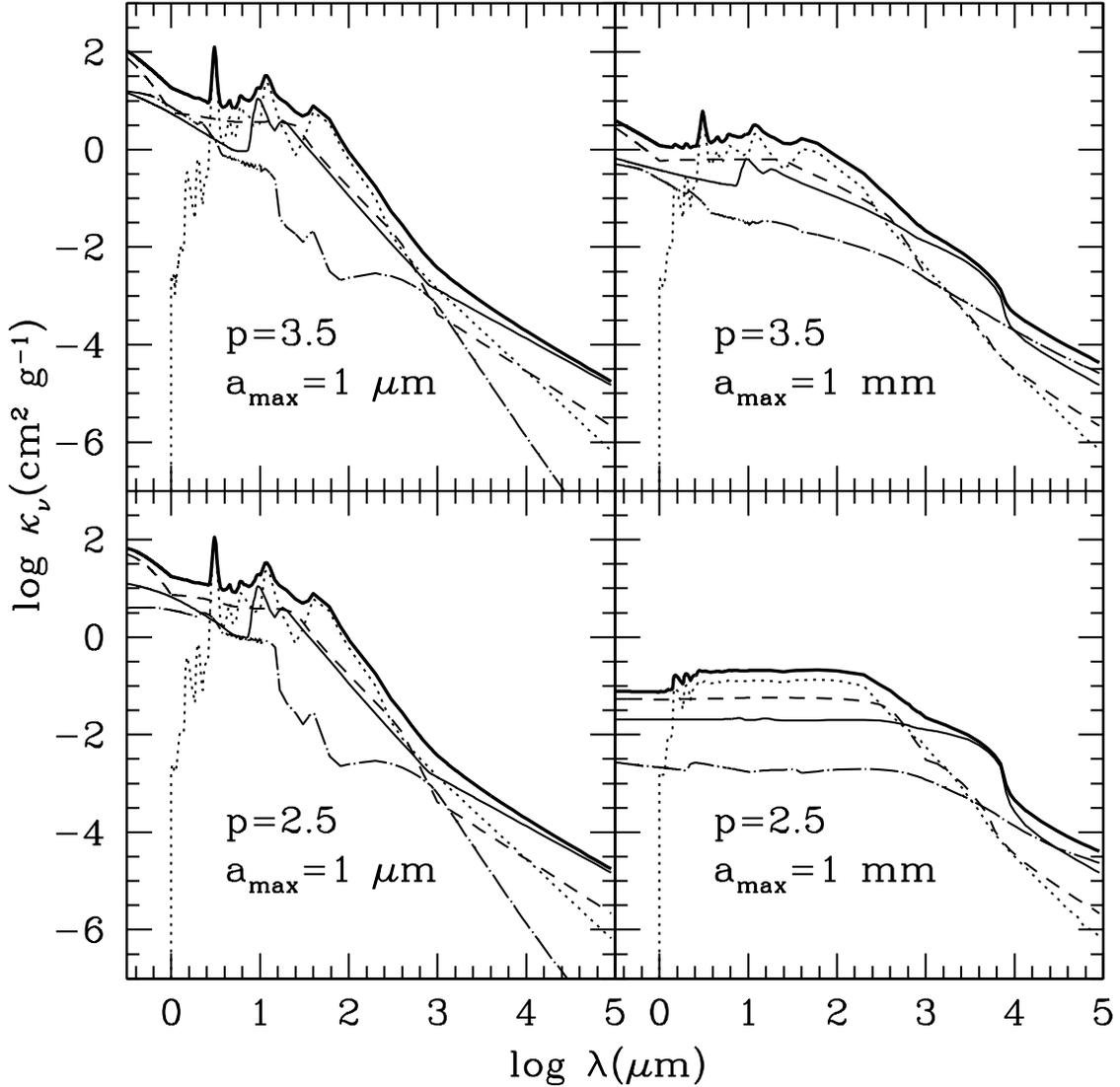}
\caption{Mass absorption coefficient $\kappa_\nu$ vs wavelength
for a grain size distribution characterized by  $p=3.5$ (upper panels)
and $p=2.5$ (lower panels), with $a_{max}=$ 1 $\mu$m (left panels) 
and  1 mm (right panels), for a temperature $T=$ 100 K.
Each curve  corresponds to a different kind of grain: silicates 
(light solid line), 
water ice (dotted line), organics (dashed line), troilite (dot-dashed line). 
The total mass absorption coefficient is represented with a solid thick line.
Each panel is labeled 
by $p$ and $a_{max}$.}
\label{fig_opa_ingredients}
\end{figure}

Figure \ref{fig_opa_ingredients} shows the contribution of the 
different grains to 
the absorption coefficient for $T=$ 100 K,  $a_{max}=$1 $\mu$m 
and 1  mm and $p=$ 2.5 and 3.5. 
In general, the relative contribution of each type of grain depends on the 
wavelength and size distribution.
For $a_{max}=$ 1 mm, silicates are the most important contributor at 
$\lambda \sim$ 1 mm, and troilite becomes of comparable importance 
in the cm range. 
In the mid- and far-IR range, water ice dominates, and the second important 
contributor are organics. 
For $a_{max}=1 \ \mu$m, silicates dominates the opacity in the mm and 
cm range, and troilite is not very important. In the mid- and far-IR water ice 
is dominant, with important contributions of silicates and organics. 
In particular, the organics tend to increase the opacity at 
$\lambda \lesssim 6 \ \mu$m, decreasing the contrast between the 
silicate band at 10 $\mu$m, and the adjacent continuum, respect 
to opacity calculations 
considering only silicates.    
The silicate bands at $\sim 10-20 \ \mu$m almost disappear from 
the total absorption coefficient for $a_{max} =$ 1 mm (see also 
Figure \ref{fig_opa_varias}), even for temperatures higher than the 
ice sublimation temperature. 
We find a similar result for $a_{max} \gtrsim 10 \ \mu m$ (
in agreement with P94).

\begin{figure}
\plotone{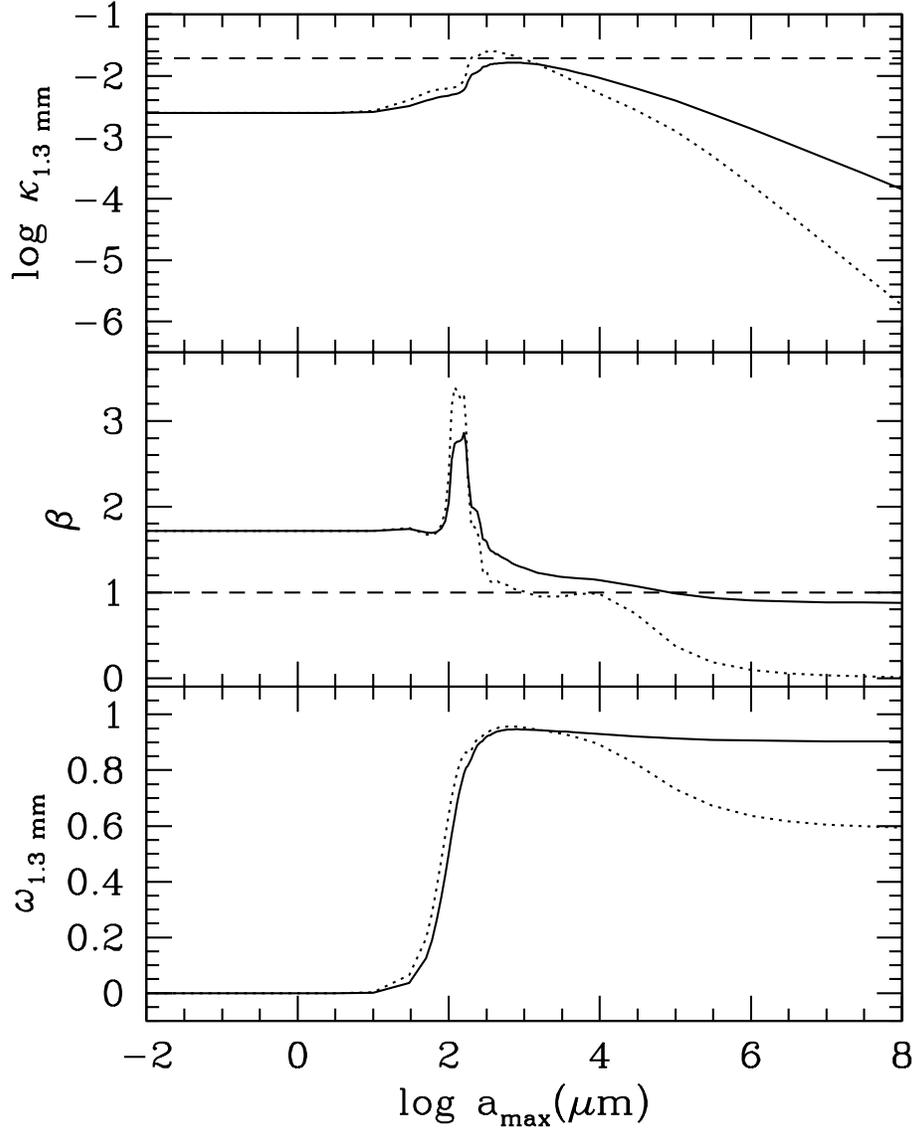}
\caption{Upper Panel: Mass absorption coefficient at $\lambda=$  1.3 mm
as a function of the maximum grain radius
$a_{max}$.
The solid line corresponds to $p=3.5$ and the dotted line to $p=2.5$.
The horizontal dashed line represents the
frequently adopted opacity at 1.3 mm
from BS91. 
Middle panel: $\beta= {\rm d \log} (\kappa_\nu) / {\rm d \log} (\nu)$,
calculated between $\lambda=$ 0.769 and 1.3 mm
for the same cases as the upper panel.
The horizontal dashed line is $\beta=$1. 
Lower panel: albedo at 1.3 mm for the same cases shown in the upper panel.}
\label{fig_opa_mm}
\end{figure}

Figure \ref{fig_opa_mm} (upper panel) shows the 
mass absorption coefficient at $\lambda= 1.3$ mm as a function of
maximum grain radius $a_{max}$ for $p=$ 2.5 and 3.5, and for a 
temperature $T < 100$ K.
The opacity is almost constant
for $a_{max} \le$ 10  $\mu$m, has a maximum around $a_{max} \sim$ 1  mm and
decreases for larger $a_{max}$ as $\kappa_\nu \propto a_{max}^{3-p}$ 
for 3 $< p <$ 4  and
$\kappa_\nu \propto a_{max}^{-1}$ for $p < 3$.
This happens because
the absorption coefficient at $\lambda \sim$ 1  mm is dominated
by grains with $a \sim $1 mm and the mass of those grains is maximum 
for $a_{max} \sim$ 1 mm and
decreases for larger $a_{max}$, with a slope that depends on $p$.
Our results agree closely with those of MN93, even though we use a 
different dust composition. 

The straight line in Figure \ref{fig_opa_mm} (upper panel) is the 
opacity at 1.3 mm
derived from the formula
$ \kappa_\nu = 0.1 (\lambda/250 \ \mu m)^{-\beta}$, with $\beta=1$ 
(BS91),
commonly used to derive disk masses from mm-wave fluxes.
At $\lambda=$1.3 mm the value of this absorption coefficient is close to the
maximum absorption coefficient, achieved when $a_{max} \sim $ 1 mm,
but is generally much larger for other grain sizes. 
In addition,
according to P94, composite, unfilled grains have 
in general smaller opacities at 1 mm than compact spheres of the same size.

The frequency dependence of the dust opacity in the wavelength regime of 
interest, 
parameterized by $\beta = d\log \kappa_{\nu}/d\log \nu$,
can in principle be used to constrain grain growth (BS91; MN93), depending 
on the
chemical composition (P94).  In Figure \ref{fig_opa_mm} (middle panel), we 
show the value of 
$\beta$ in the mm-wave range,
\be
\beta= {\log(\kappa_{0.769 {\rm mm}}) - \log(\kappa_{1.3 {\rm mm}}) 
\over \log(1.3 {\rm mm})-\log(0.769 {\rm mm})},
\en
as a function of maximum grain radius $a_{max}$ for the same cases
plotted in the upper panel.
For small grains ($a_{max} \le 10  \ \mu$m), $\beta =$  constant $\sim$ 1.75,
and approaches a maximum value for sizes $a_{max} \sim 60- 300 \ \mu m$ 
(corresponding
to $2 \pi a/\lambda \sim 1$).  The value of $\beta$ decreases with 
increasing maximum size,
and for $a_{max} >$ 10 cm, becomes independent of $a_{max}$.
Our results are qualitatively consistent with those of MN93.

The detailed behavior of $\beta$ for $a_{max}$ between 100 $\mu m$ and 1~cm
depends on the relative importance of the
different kind of grains. Small changes in abundances and optical
properties can change the value of $\beta$ in these
intermediate sizes.
In particular, according to P94, composite grains with 50 \% void
volume have a less steep increase of $\beta$ with $a_{max}$, for grain sizes
smaller than 1 mm.

Figure \ref{fig_opa_mm} (lower panel) shows that the  albedo $\alb$
at $\lambda=1.3$ mm
increases from 0 to $\sim$ 0.9  for $a_{max}$ between 10 $\mu$m and 1 cm 
(MN93).
For larger values
of $a_{max}$,   $\alb \sim$ 0.9 for p=3.5 and $\alb \sim$ 0.5 for $p=$2.5.
This large albedo means that scattered light contributions in the mm-wavelength
range may not be neglected (MN93).

\section{Disk Models}
\label{sec_model}

We solve self-consistently the complete set of vertical structure
equations, including irradiation and viscous heating, resulting
in detailed profiles of temperature and density with
vertical height and disk radius.  
The assumptions, equations and method are 
described in Papers I and II.  
In particular, we assume that gas and dust are well mixed and 
in thermal balance everywhere.
The input parameters to calculate the disk structure  are the disk mass 
accretion rate $\Mdot$, the viscosity parameter $\alpha$, and 
the central star mass $M_*$, radius $R_*$ and temperature $T_*$.
Whenever it is possible, we use observational constraints to reduce the 
parameter space that must be explored. 
To this end we have chosen typical parameters of T Tauri 
stars from  \cite{gull98}: 
$\Mdot=3 \times 10^{-8} \MSUNYR$, $M_*=0.5 \ \MSUN$, $R_*=2 \ \RSUN$ 
and $T_*=4000$ K
for the fiducial model.  We assume the disk is in steady state accretion, i.e.
the accretion rate through the outer disk is the same as in the inner disk,
at a rate inferred from measurement of the ultraviolet and optical 
continuum excess emission produced 
as material accretes via a shock at the stellar surface
(\cite{CG98}).
The viscosity parameter $\alpha$  is changed appropriately 
to fix the disk mass.
The adopted values of $\alpha$ are between 0.005 and 0.036,
corresponding to disk masses between 0.1 and 0.02 $\MSUN$ 
(for $\Mdot=3 \times 10^{-8} \ \MSUNYR$ and $R_d=$ 100 AU).  
The disk outer radius is taken to be $R_d=$ 50, 100 and 200 AU, 
and the inner radius is $R_{hole}=3 \ R_*$.

Given the disk structure,  the emergent intensity is calculated by 
integrating the transfer equation,
\be
I_\nu = I_\nu^{bg} \ e^{-\tau_{\nu,i}}+\int_0^{\tau_{\nu,i}} S_\nu (t_{\nu,i}) \  e^{-t_{\nu,i}} \ dt_{\nu,i} \ ,
\label{eq_transf}
\en
along rays that pierce the disk. Here,
$S_\nu$ is the source function,  
$\tau_{\nu,i}$ is the total optical depth along a particular ray 
(with an inclination angle $i$ respect to the symmetry axis of the disk),
$t_{\nu,i}$ is the variable optical depth along the ray, and $I_\nu^{bg}$ 
is the background radiation, i.e.,
$I_\nu^{bg}=B_\nu(T_{bg})$ with $T_{bg}=2.73$ K.
The optical depth is calculated including true absorptions and scattering 
events.

Since the dust albedo of large grains is so large in the mm range 
(Figure \ref{fig_opa_mm} and Table 1,
it is important to include the contribution of the 
emissivity in scattered light. 
Assuming coherent and isotropic scattering, the source function is  
taken to be (\cite{M78})
\be
S_\nu =(1-\alb) B_\nu + \alb J_\nu \ ,
\en
where $B_\nu$ is the Planck function and $J_\nu$ is the mean intensity 
of the radiation field.
We approximate the 
$J_\nu$ by the mean intensity of a vertically isothermal slab  (see
MN93), with the Planck function evaluated at the local temperature,
\be
J_\nu(t_{\nu,0}) = B_\nu(t_{\nu,0}) \biggl \{ 1 + {\exp{[-\sqrt{3(1-\alb)} \ t_{\nu,0}]} 
+ \exp{[\sqrt{3(1-\alb)} \ (t_{\nu,0}-\tau_{\nu,0})]}
\over \exp{[-\sqrt{3(1-\alb)} \ \tau_{\nu,0}]} ( \sqrt{1-\alb} -1) -(\sqrt{1-\alb}+1)} \biggr \} \ \ ,
\label{eq_jota}
\en
where $\tau_{\nu,0}$ is the total optical depth in the vertical direction 
($i=0^\circ$) and $t_{\nu,0}$
is the variable optical depth in the same direction.

In an attempt to simulate the observational procedure of 
taking on-off measurements, 
the background intensity $I_{bg}$ 
due to the cosmic microwave background 
is subtracted 
from the disk emergent intensity given by equation (\ref{eq_transf}). 
The observed flux is calculated by integrating the 
intensity, emerging in a direction defined by the line of sight, over 
the solid angle subtended by the 
disk as seen by the observer.
For optically thin regions, the emergent flux calculated including scattering 
is similar to the flux calculated assuming the dust is purely absorbing. 
For optically thick regions, with  $\tau_{\nu,0} >> 1/(1-\alb)$, the 
flux for a pole-on disk  is smaller by a factor roughly 
$1-\alb/[\sqrt{1-\alb}+1][\sqrt{3(1-\alb)}+1]$  
 than the purely absorbing case, 
consistent with the discussion in MN93. For intermediate optical depths 
(e.g. $\tau_{\nu,0} \sim$ 1), the behavior of the emergent flux cannot be 
described by 
analytic expressions.

\section{Results}
\label{sec_results}

\subsection{Disk ``photosphere'' and hidden central stars} 
\label{sec_surface}

Figure \ref{fig_opa_varias} shows that 
grain growth affects the opacity at both small and large wavelengths.  
For a fixed mass of
dust, increasing $a_{max}$ tends to increase the mm-wave opacity 
(for $a_{max} \lesssim mm$) 
at the same time as
it decreases the
 opacity at the characteristic wavelength of the stellar radiation,
$\lambda \sim$ 1 $ \mu$m.  This has important effects on the disk 
temperature distribution
and thus the disk structure.

\begin{figure}
\plotone{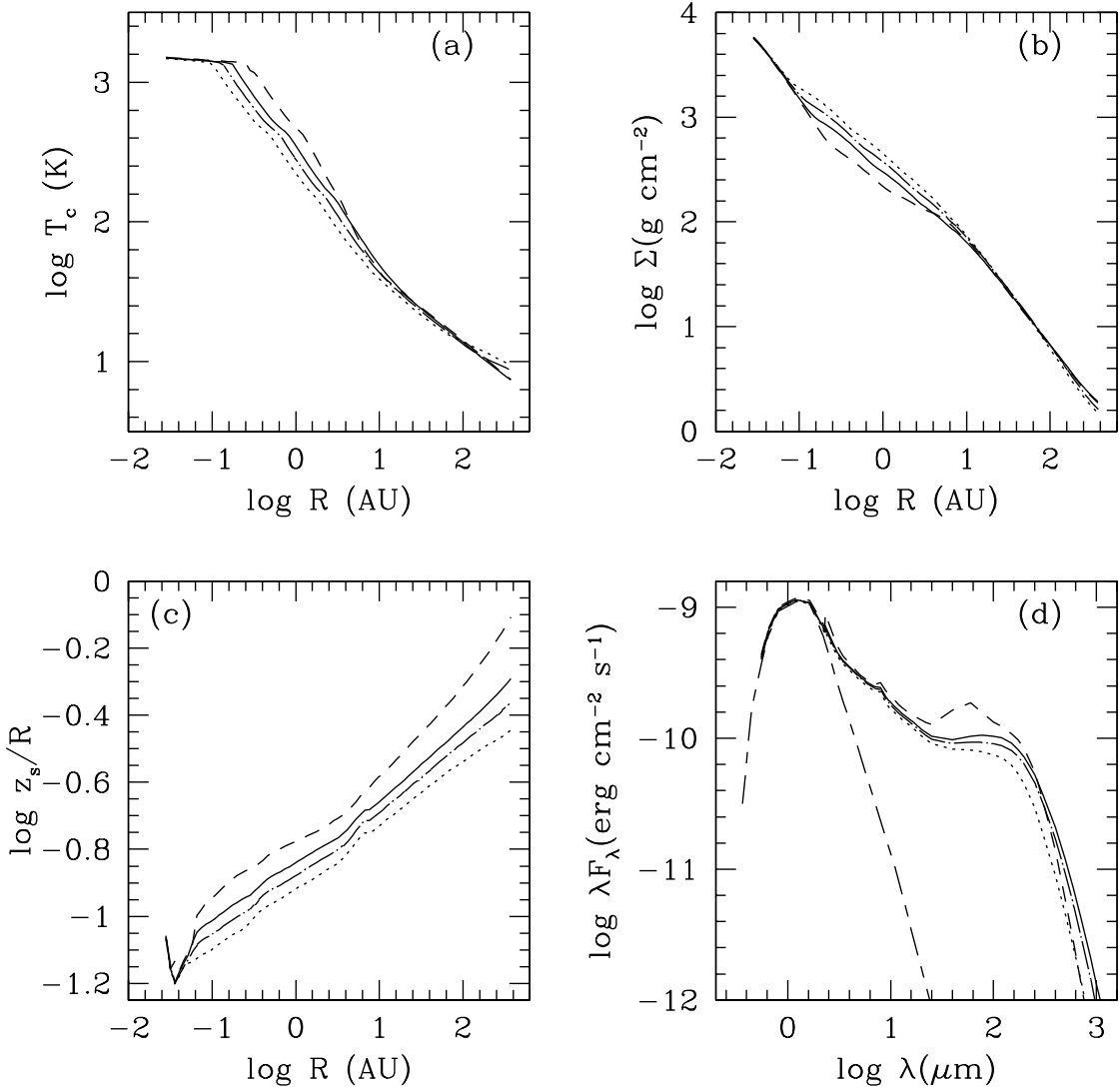}
\caption{Disk structure and SED for different maximum grain 
size and exponent p=3.5.
Each curve corresponds to a different value of maximum grain 
radius,   $a_{max}=$ 10 $\mu m$ (dashed line), 1 mm (solid line), 1 cm 
(dot-dashed line) and 10 cm (dotted line). 
Each panel represents the radial distribution of: 
(a) Midplane temperature; 
(b) Mass surface density; (c) Height of the irradiation surface divided 
by radius. In panel (d) the resulting SEDs are shown, 
assuming the disks are oriented pole-on. The SED of the central star is 
also plotted (long-short dashed line). }
\label{fig_estruc_p3p5} 
\end{figure}

\begin{figure}
\plotone{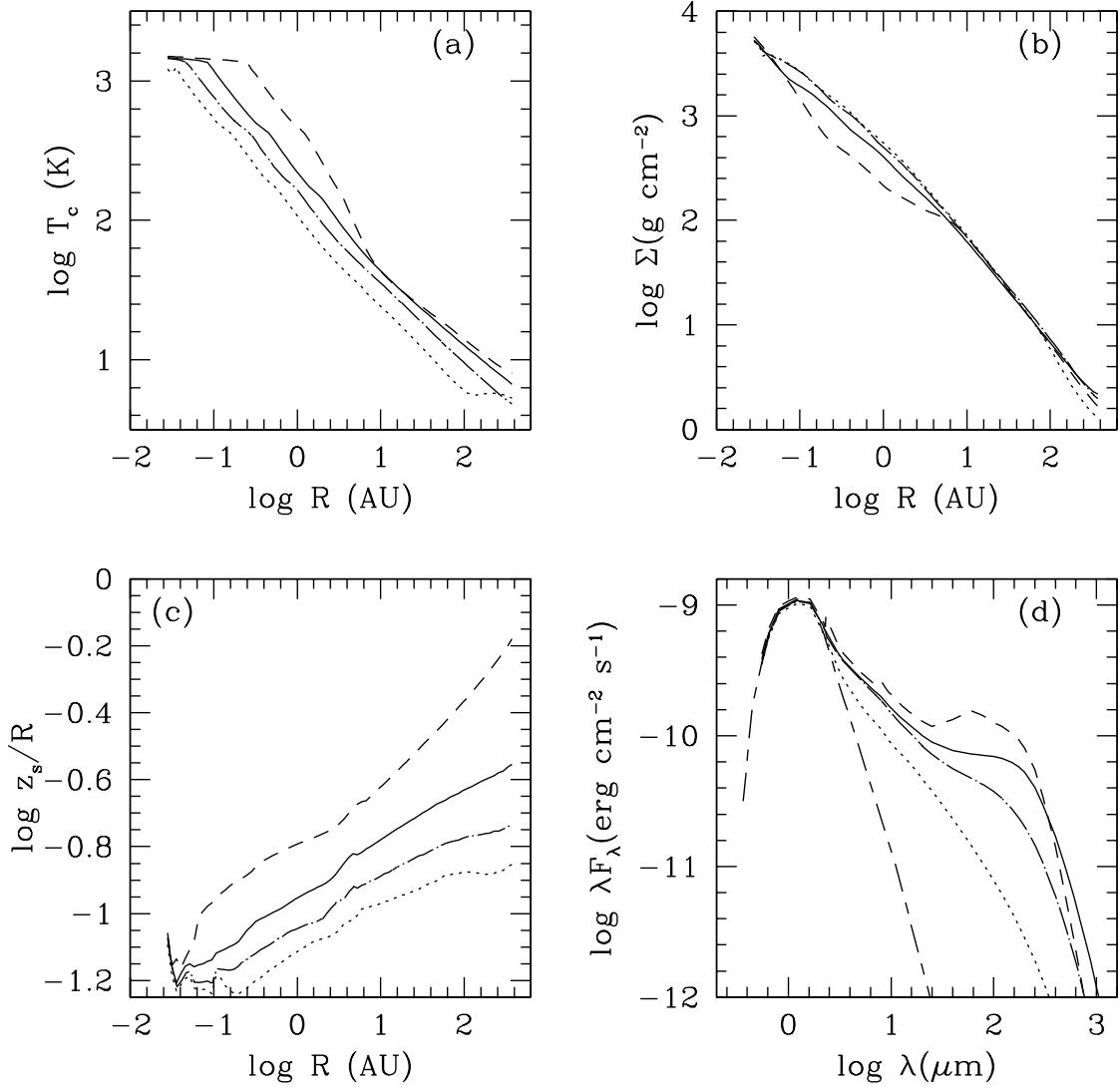}
\caption{Disk structure and SED for different maximum grain 
size and exponent p=2.5.
See caption of Figure \ref{fig_estruc_p3p5}.}
\label{fig_estruc_p2p5} 
\end{figure}

Figures \ref{fig_estruc_p3p5} and \ref{fig_estruc_p2p5}  show 
the midplane temperature $T_c$, surface density $\Sigma$ and  the 
height of the ``irradiation surface'' $z_s$ as a function of 
radius, and the SEDs of pole-on ($i=0^\circ$) disks, 
 for different grain size distributions and the same 
disk mass $M_d \approx 0.046 \ \MSUN$, accretion rate 
$\Mdot=3 \times 10^{-8} \ \MSUNYR$ and the typical central star 
properties mentioned in \S \ref{sec_model}.
The height $z_s$ is defined to be where the mean 
optical depth to the stellar radiation is 
unity (using the mean opacity $\chi_P^*$, listed in Table 1) 
and represents where most of the 
stellar radiative energy is deposited (see paper II).
The height of this surface decreases 
when the grain sizes increases. 
Thus, the fraction of stellar radiation intercepted by the disk 
decreases, resulting in a colder disk 
than what would be expected for 
smaller grains. 
The mass surface density of the outer disk is $\Sigma \sim 1/R$ 
for $R>$ 10 AU, thus the 
disk mass is dominated by the contribution of the outer radii. 
Since the present models have all the same disk mass, they also 
have the same surface density for $R \gtrsim$ 10 AU. 

\begin{figure}
\plotone{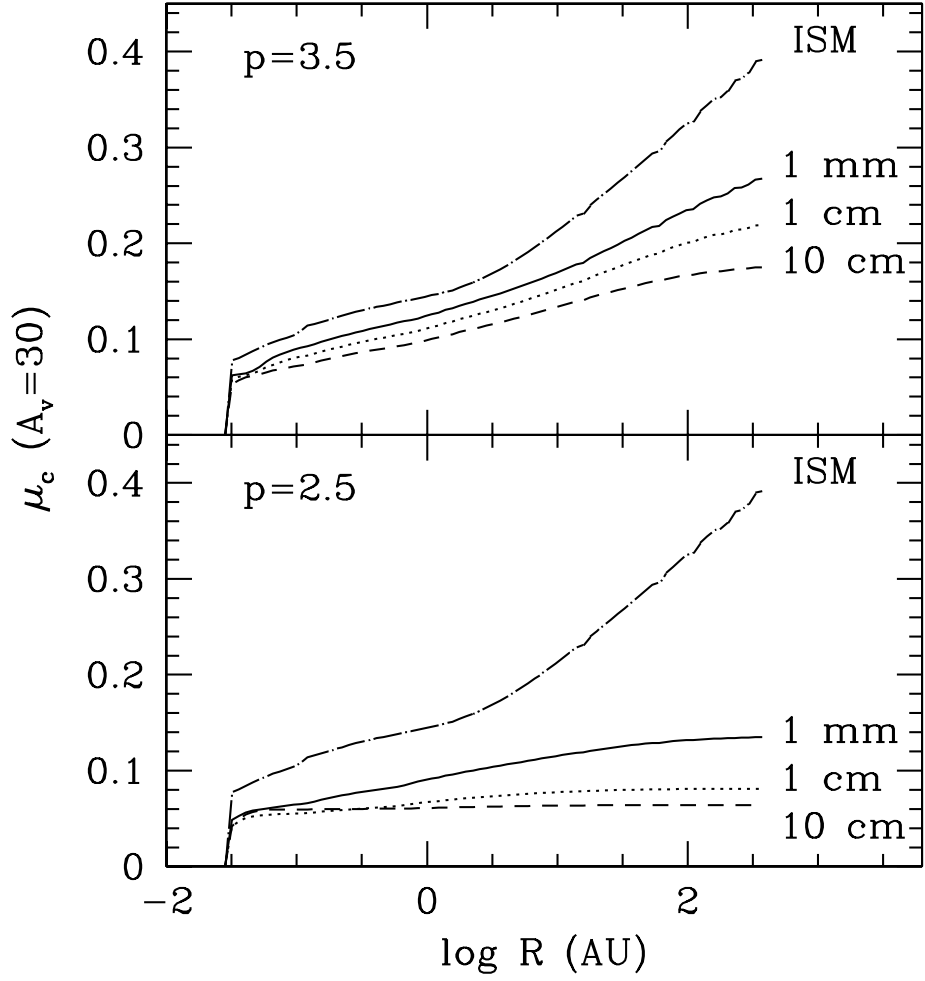}
\caption{Cosine of critical inclination angle for disks models with 
$M_d=0.046 \ \MSUN$ ($\Mdot=3 \times 10^{-8} \ \MSUNYR$, $\alpha=0.01$). 
The surface is defined so the extinction of the central star produced 
by the disk is $A_V=30$, for 
dust abundances and optical properties described in \S \ref{sec_opa} and
different distributions of grain sizes: $a_{max}=$ 1 mm (solid line), 
1 cm (dotted line) and 10 cm (dashed line). The upper panel corresponds 
to $p=3.5$ and the lower panel,  to $p=2.5$. For comparison we also 
show $\mu_c$  for ISM-dust, with ingredients and optical properties 
from \cite{DL84}, $p=3.5$, $a_{max}=0.25 \ \mu m$ (dot-dashed line). }
\label{fig_surface}
\end{figure}

In Paper II we showed that if the disk has ISM-dust well-mixed with the gas, 
then
a large fraction of the CTTS should have their central star extinguished 
with $A_V > 30$ mag 
(30 \% of all objects with an outer disk radius $R_d=$ 100 AU).
Also, the two-layer models of Chiang \& Goldreich (1997,1998) implies 
a similar large fraction of highly-extinguished central stars. 
As discussed in Paper II, this fraction is too high in comparison with 
estimates from
current observational surveys in the Taurus molecular cloud complex.
The decrease in the optical and near-infrared dust opacity with 
increasing $a_{max}$ results
in (optically and geometrically) 
thinner disks, and thus the fraction of highly-extinguished stars 
predicted for a random
distribution of viewing inclinations decreases. 
Figure \ref{fig_surface} shows the cosine of the critical angle $i_c$, 
such that $\cos \ i_c= \mu_c= z_{A_V}(R)/[z_{A_V}(R)^2+R^2]^{1/2}$, 
where $z_{A_V}(R)$ is the height at which the the extinction coefficient 
is $A_V= 30 $. 
Assuming a random distribution of disk rotation axes to the line of sight,
for $a_{max}=$ 1 mm and $p=3.5$, 20 \% of the T Tauri 
stars with disks with a typical radius $R_d \sim$ 100 AU, 
should be extincted by their
disks with an $A_V > 30$ mag, while for $a_{max}=$ 10 cm and $p=2.5$ only 
$\sim 6$ \% of T Tauri should have $A_V > 30$ mag.
Because the fraction of such disk-obscured sources in Taurus is 
estimated to be not more than
about 15 \% (Paper II), it is clear that the grain-growth models 
are in much better agreement
with the current observational constraints.

The SEDs of the models show that the larger the 
fraction of big grains, the smaller the mid- and far-IR fluxes. On the other 
hand, the mm fluxes increases with $a_{max}$ until $a_{max} \sim$ 1 mm, and 
decreases with $a_{max}$ for larger grains, reflecting 
the behavior of  the opacity  shown in Figure \ref{fig_opa_mm}.  
In the next section we show how the SED of a disk model with a 
larger fraction of big grains than the ISM dust compare to 
observations.

\subsection{Median SED}
\label{sec_median}

\begin{figure}
\plotone{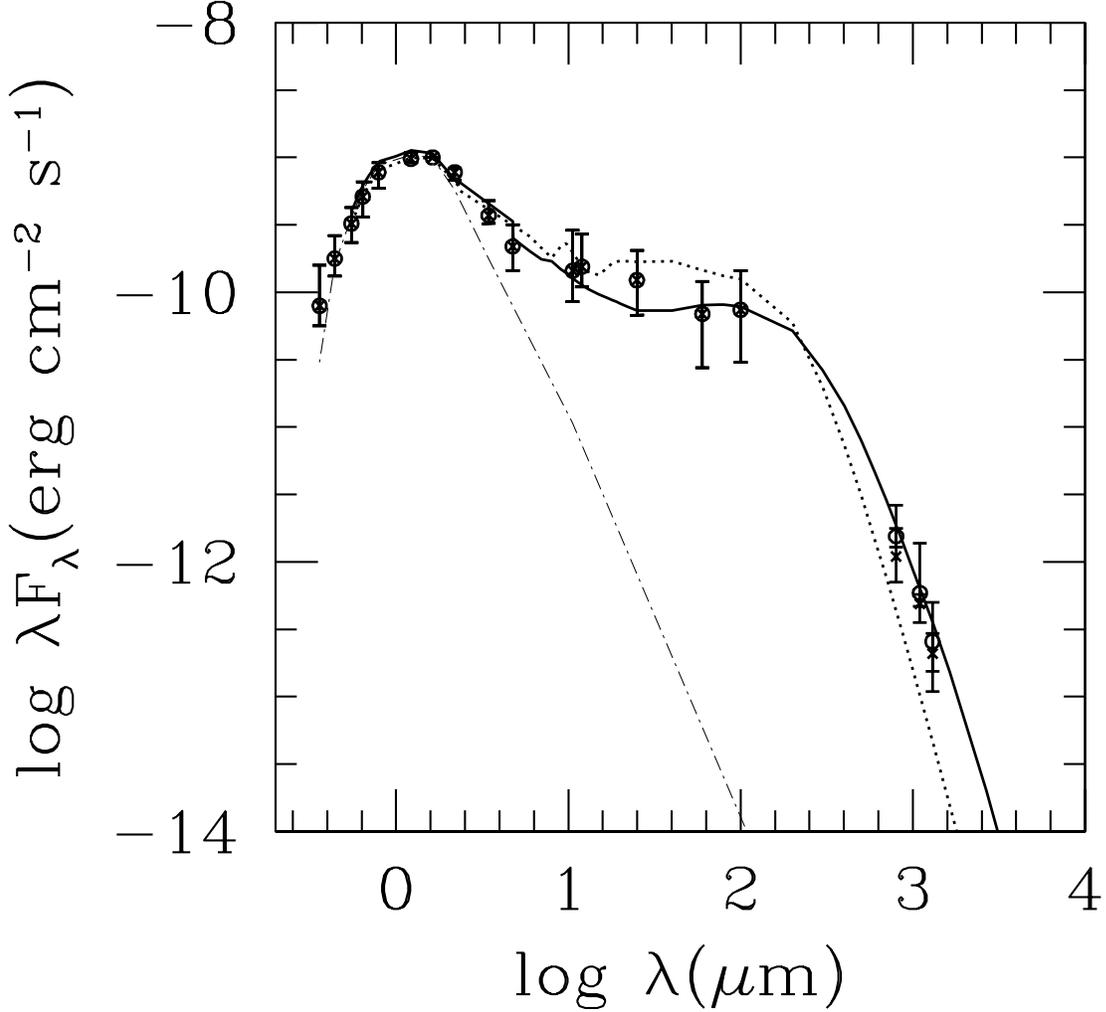}
\caption{SED of the model with $M_d=0.046 \ \MSUN$, 
$\Mdot=3 \times 10^{-8} \ \MSUNYR$, $\cos i=$ 0.65, $R_d=$ 100 AU, for 
ISM-dust (dotted line) and for the dust model adopted in this paper, 
with  $p=3.5$ and $a_{max}=$ 1 mm (solid line). The points represent 
the median observed SED and the errorbars are the quartiles. The 
observed fluxes are normalized at $\lambda=1.6 \ \mu m$. For the mm 
wavelengths we also show the median fluxes without normalization 
(crosses)  (normalization is a reasonable procedure for optically 
thick emission, but since the mm flux could have an important 
contribution from optically thin region, the true mm median 
would be something between the normalized and the non 
normalized fluxes). The contribution of the UV and optical 
excess emission produced 
	in an accretion shock is not included in the model SEDs.}
\label{fig_median}
\end{figure}

Figure \ref{fig_median} shows the spectral energy distributions 
(SEDs) for two disk models, both with 
$M_d=0.046 \ \MSUN$ ($\Mdot=3 \times 10^{-8} \ \MSUNYR$, $\alpha=0.01$, 
$R_d=$ 100 AU) and a central star with $M_*=0.5 \ \MSUN$, $R_*=2 \ \RSUN$ and 
$T_*=$4000 K  ($L_*=0.9 \ \LSUN$), and an inclination angle $\cos \ i=0.65$
\footnote{We use $\cos \ i=0.65$ instead of 0.5 for the average inclination to account for
the occultation in edge-on systems, which would drop out of the observational
samples due to weak optical and infrared emission.}. 
We adopt the distance to Taurus of 140 pc (\cite{KDH94}).
One of the models has ISM-dust and the second model has dust with the abundances 
and optical properties described in \S \ref{sec_opa}, and $a_{max}=$ 1 mm, $p=$3.5. 
The main difference between both models are that the ISM-dust disk model has larger 
far-IR and smaller mm-wave fluxes than the 
grain growth model. Both models are compared 
to the median SED for the T Tauri stars in the Taurus-Auriga molecular cloud (
from \cite{KH95}, c.f., Paper II).  
The SED for the $a_{max}=$ 1 mm disk model is much closer to the observed median spectrum.
A range of values of $a_{max}$ and $p=$ 2.5 - 3.5 produce 
SEDs consistent with the quartiles (c.f. Figures  \ref{fig_estruc_p3p5} 
and \ref{fig_estruc_p2p5}). Also, disks must have different radii, inclination angles, 
environment, etc., which would contribute to the scatter in the observed SEDs (c.f., paper II).

\subsection{Scattered light images}
\label{sec_images}

\begin{figure}
\plotone{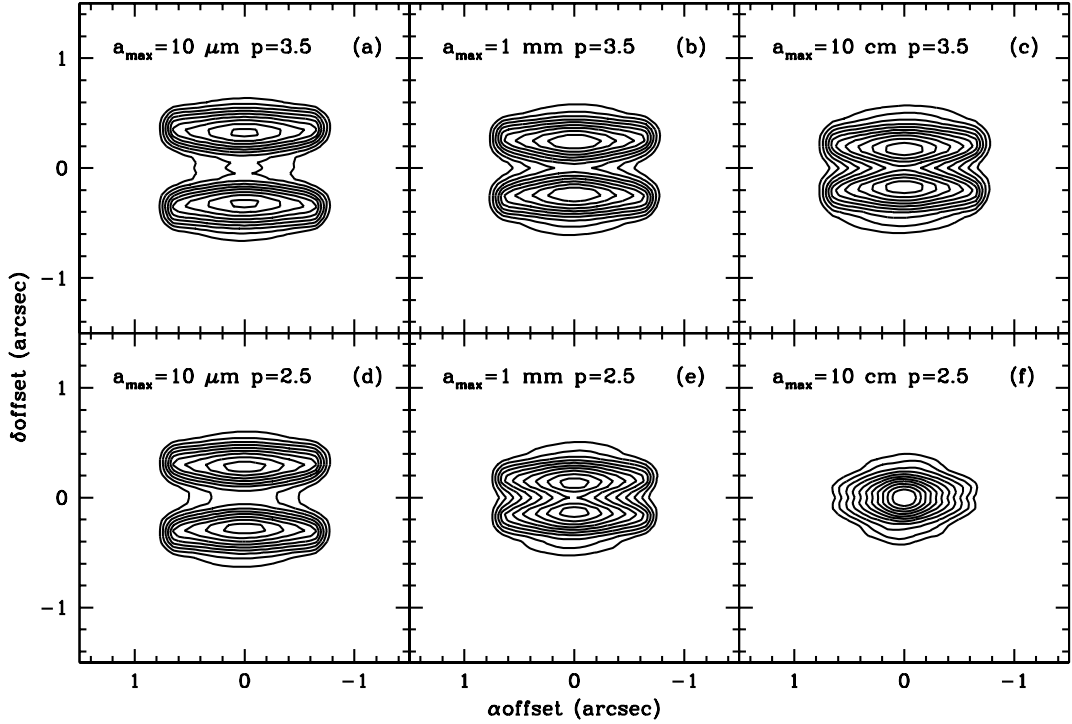}
\caption{ Images of edge-on disks at $\lambda=0.814 \ \mu m$ 
for dust mixtures with p=3.5 (upper panels) and
p=2.5 (lower panels), and $a_{max} = 10 \ \mu$m (left),
$a_{max} = 1 $mm (center), and $a_{max} = 10 $cm (right). 
The model has $M_d=0.046 \ \MSUN$, and a central star with 
$L_*=0.9 \ \LSUN$ and 
$T_*=$ 4000 K.
 Each panel is labeled with the corresponding
$p$ and $a_{max}$.  The images are calculated at $\lambda=0.814 \ \mu m$
and convolved with the PSF of the HST PC1. 
The outermost contour level corresponds to 18.3 mag arcsec$^{-2}$, 
the peak contour level is 13.3 mag arcsec$^{-2}$, and the contour 
interval is 0.5 mag arcsec$^{-2}$ }
\label{fig_images}
\end{figure}

The decrease in opacity at small wavelengths when the fraction of large grains 
increases also affects images in stellar scattered light. 
Figure \ref{fig_images} shows images in scattered light 
of edge-on disk models ($i=90^\circ$), with the typical properties 
described in \S \ref{sec_model} and a viscosity parameter $\alpha=0.01$. 
These disk models have $R_d=$ 100 AU and $M_d=0.046 \ \MSUN$, and again
are assumed to lie at a distance of 140 pc.
The images are calculated at $\lambda=0.814 \ \mu m$, 
in the single scattering approximation,  
assuming a Henyey-Greenstein phase function, with an asymmetry parameter 
$g=0.65$. 
The emergent intensity distribution has been convolved with the HST PSF 
(for the filter F814W/PC1). 
The aspect ratio of the
image, i.e. the width of the central dark lane or
equivalently the distance between
emission maxima, is very sensitive to $a_{max}$ for a given $p$. 
In this case, the larger the value of $a_{max}$ 
 the smaller the width of the dark lane,
 which almost disappears for $p=2.5$, $a_{max}=$10  cm. 
 This occurs because 
 the opacity at $\lambda \sim 0.8 \ \mu m$ decreases when $a_{max}$ increases
 and $p$ decreases (c.f \S \ref{sec_opa} and Table 1), and both
 the height of the  surface where
 the stellar radiation is attenuated (c.f \S \ref{sec_surface}) and
 the extinction of the scattered light at a given height produced by the 
outer disk decrease.
 For the same reason, the maximum intensity of the images increases 
with $a_{max}$. 
 The width of the dark lane also depends on the outer disk radius.
 The height where most of the stellar radiation is deposited
($z_s$, c.f. Figures  \ref{fig_estruc_p3p5} and \ref{fig_estruc_p2p5}) 
evaluated at the the disk
radius can be used to make a rough 
prediction of the  distance between the emission peaks.
It is not very sensitive to the disk inclination,
as long as the the disk occults the central star.  

In Paper II we showed that the near-IR scattered light image 
predicted for the well-mixed ISM-dust models produces an
aspect ratio $\sim 2$ times larger than is 
consistent with {\em Hubble Space
Telescope} images of HK Tau/c and HH 30.  The results shown in 
figure \ref{fig_images}
show that grain growth will improve the comparison between theory and 
observation.
Here we consider a detailed comparison between the the image of HK Tau/c 
and a disk model.

\begin{figure}
\plotone{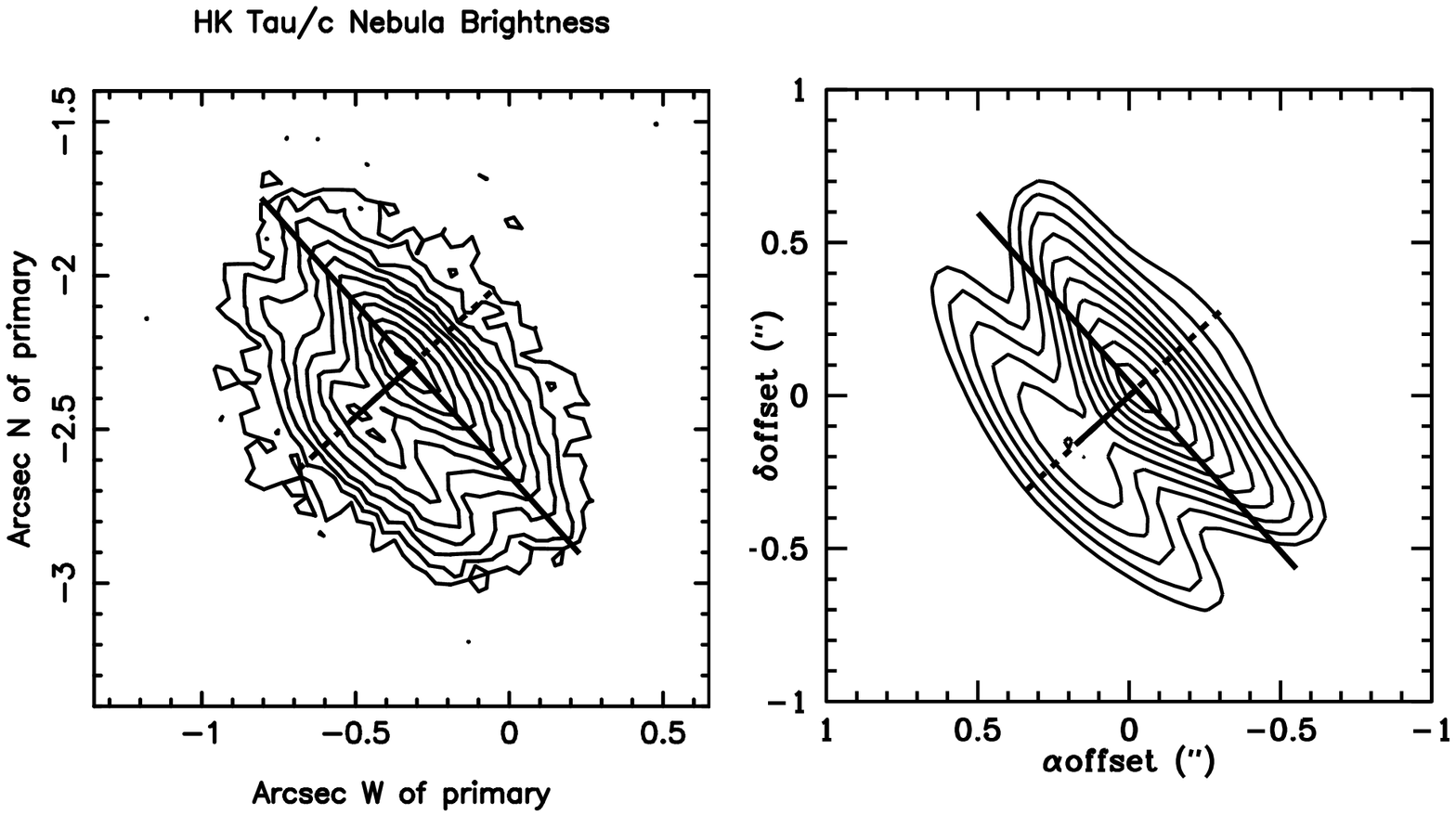}
\caption{The left panel is the image of HK Tau/c (\cite{sta98}) 
and the right panel is the image of a disk model with $\Mdot=3 \times 10^{-8} \ \MSUNYR$, 
$\alpha=0.01$, $R_d=110$ AU, $M_d=0.065 \ \MSUN$, $M_*=0.5 \ \MSUN$, 
$R_*=1.3 \ \RSUN$, $T_*=$3500 K, $a_{max}=$ 1 m, $p=3.5$. 
The disk inclination angle $i=80.8^\circ$ and the P.A. is $40 ^\circ$. 
 The levels are the same than in Figure \ref{fig_images} 
and in the observed image.
}
\label{fig_hktau}
\end{figure}

Figure \ref{fig_hktau} shows the image of a disk model with  
$\Mdot=3 \times 10^{-8}$, $\alpha=0.01$, $T_*=3500$ K, 
$R_*=1.3 \ \RSUN$ ($L_*=0.22 \ \LSUN$), 
$M_*=0.5 \ \MSUN$, $R_d=110$ AU, inclination angle $i=89.8^\circ$ 
and position angle P.A.$=40^\circ$ . The disk model dust is characterized by 
$a_{max}=1$ m and $p=3.5$, and the total opacity and albedo are given 
in Table 1.
 
The model image is compared to the image of HK Tau/c (\cite{sta98}), with 
the same contour levels in both images as in Figure \ref{fig_images}.
The observed image and the model show reasonable agreement,
as for instance the distance between maxima, the aspect ratio and the 
maximum brightness. 
The flux at $\lambda=$1.3  mm predicted by this model is $F_{1.3 \ mm}=$ 
8 mJy, 
consistently
smaller than the flux for the unresolved HK Tau binary system reported by 
\cite{BSCG} (= BSCG), $F_{1.3 \ mm} =$ 41 mJy.
The mass of this disk model is $M_d=0.065 \ \MSUN$.
We find that growth to a very large value of $a_{max}$ yields a 
dramatically reduced
mm-wave opacity, and thus a very low mm-wave flux for this large disk 
mass (see \S 5),
while producing an optical image in good agreement with observations.

On the other hand, dust grains with $a_{max}=$ 1 mm and $p=2.5$ 
have, approximately, the same opacity and albedo at $\lambda=0.8 \ \mu m$, 
but a much larger opacity at 1.3 mm (c.f., Table 1). 
We made a disk model with the same mass and central star properties 
than the one shown in 
Figure \ref{fig_hktau} and we found it 
has a similar scattered light image, but 
a much larger flux at 1.3 mm,  $F_{1.3 mm} \sim$ 30 mJy. 
However, this flux seems too high, considering that
the total flux should have
contributions from both disks in the binary, and that the less
inclined disk of the primary is expected to have a larger
contribution.

The model shown in Figure \ref{fig_hktau} is not unique,
being under-constrained by observations. Nonetheless,
this example illustrates
how disk masses inferred from optical imaging (or even mm-wave fluxes) 
may be substantially
under-estimated.  Without additional observational constraints on dust 
properties, 
detailed conclusions are difficult to derive.

Moreover, the mass of the present model is  several hundred times larger 
than  the mass inferred by Stapelfeldt et al. (1998) 
(i.e., their models C and D have masses  
$M_d \sim 5.9 \times 10^{-5}-1.2 \times 10^{-4} \ \MSUN$).
The discrepancy is due to the difference in dust opacity.
Stapelfeldt \etal used an opacity typical of ISM dust grains, which is 
much larger at $\lambda=0.8 \ \mu m$ than the opacity corresponding to 
a larger value of $a_{max}$ (see Table 1). This implies that 
they require a much smaller disk mass than we do  
to reproduce the width of the dark lane of the image and 
its aspect ratio.  
Assuming that the disk is optically thin and 
the flux at 1.3 mm scales with disk mass and the 
absorption coefficient, using the values given in Table 1 we infer 
$F_{1.3 mm} \sim$ 0.01 mJy for their model D.
Thus, mm observations resolving the binary system 
would be very important 
to define whether this is a very low-mass disk with ISM type 
of dust or a moderate mass disk with a much smaller fraction of 
small grains than the ISM dust, as we suggest here.

\section{Disk mm-wave fluxes and masses}
\label{sec_mm}

Much of what is known or estimated about T Tauri disks comes from analyses
of mm- and sub-mm wavelength emission.  The first observational arguments
in support of grain growth were made to explain observed spectral indices
in this wavelength regime (BSCG, BS91).  Our previous models (Paper II) with
ISM-dust could not explain the observed mm-wave fluxes.
In this section we discuss our detailed model results for 
disks with larger grains in this wavelength region.

In principle, disk masses and aspects of grain properties can
be derived from long-wavelength fluxes and spectral slopes.  
The analysis is particularly easy
when the disk is optically thin.  
In this case the observed flux is proportional
to the disk mass, absorption coefficient at the wavelength of observation, 
and the source function at this wavelength.  
At long wavelengths the source
function is expected to have 
a Rayleigh-Jeans wavelength or frequency dependence,
so that the flux is then 
$F_\nu \propto  M_d \  k \  \prom{T} /c^2 \nu^2 \kappa_{\nu}$, 
where $k$ is the Boltzmann constant, $c$ is the speed of light, and 
$\prom{T}$ is a mass-weighted disk mean temperature.
The spectral index $d\log F_{\nu}/d\log \nu$ is 
the product of the intrinsic dependence of the dust opacity times the
Rayleigh-Jeans distribution.  Unfortunately, optical depth effects cannot
be ignored in general and the source function may not have a Rayleigh-Jeans
form if the dust temperature is too low (BS91), 
so in general detailed models must be 
used to compare with observations.

To make this comparison we constructed model sequences of fixed disk mass
with differing dust properties.  The stellar luminosity,
stellar mass, and radius are fixed as before at 0.9  $\LSUN$, 0.5 $\MSUN$ 
and 2 $\RSUN$ (c.f., \S \ref{sec_model}).
We assumed a fixed mass accretion rate $\Mdot = 3 \times 10^{-8} \MSUNYR$
and a fixed outer radius of $R_d = 100$~AU in our basic sequence.  Changing
the disk mass in our models then requires us to change the $\alpha$ parameter
accordingly.  Since the viscous heating in the outer disk in the basic sequence
is small, so that the temperature distribution is controlled by irradiation,
this is roughly equivalent to adopting a constant viscosity parameter 
and varying $\Mdot$ instead to change the disk mass.
 The disk masses in each sequence are $M_d= 0.023\ \MSUN, \ 0.046\ \MSUN,$ and $ 0.092 \ \MSUN$.
We assume initially that the disks are viewed pole-on, 
and discuss the effects of changing the inclination angle later.

Figure \ref{fig_flux} shows the 1.3 mm flux vs.\ $n$ for several sequences of disk models of
fixed mass (connected points)
The spectral index $n$ is given by
\be
n = {\log(F_{0.769 {\rm mm}}) - \log(F_{1.3 {\rm mm}}) \over \log(1.3 {\rm mm})-\log(0.769 {\rm mm})}.
\en
Each point along a given sequence corresponds to a different
value of $a_{max}$ (between $10 \ \mu m$ and 1 cm);
the two panels correspond to $p=3.5$ (upper) and  $p=2.5$ (lower).

We discuss the model sequences beginning on the right-hand side of the figure, where the models have
small $a_{max}$, and moving counter-clockwise along the curves as $a_{max}$ increases.  For small
$a_{max}$, the spectral index is relatively large, reflecting the intrinsic wavelength dependence
of the dust opacity, though the actual value of $n$ departs from the Rayleigh-Jeans limit
$2 + \beta \sim 3.75$ because this limit is not applicable in this wavelength range
for the low temperatures ($\sim 10-20$~K)
characteristic of the outer disk.  The flux levels scale nearly linearly with the mass, as expected in the
optically-thin case; only the inner, high-surface density regions are optically thick, and these
regions do not contain much mass.
Proceeding to larger $a_{max}$, the spectral index decreases, both because larger regions of the disk become
optically thick due to the increase in opacity (Figure \ref{fig_opa_mm}) and 
because of the decrease in $\beta$, 
which affects the frequency dependence of the emission from the optically thin regions
(cf. Figure \ref{fig_opa_mm}).
As a consequence of the grain albedo, the total optical depth of the disk has to be
$\tau_0 \gtrsim 1/(1-w)^{1/2}$ for the emergent flux to approach the optically thick limit
for pure thermal emission. 
For the models plotted in Figure \ref{fig_flux}, this limit is reached only by the most massive disk 
($M_d \sim 0.1 \ \MSUN$)  when $a_{max}=$ 1 mm.  Finally, for much larger $a_{max}$, the disk models
become mostly optically thin, and so the fluxes and spectral indexes much more closely reflect
the dust opacity and disk mass.
The $p = 2.5$ models show much more variation in $n$ at large
dust sizes because $\beta$ changes more rapidly with increasing $a_{max}$ than in the $p=3.5$ case
(Figure \ref{fig_opa_mm}).  For reference we also show the flux and spectral index 
of disk models with the same masses calculated assuming they contain ISM-dust (large open triangles). 
In this case the outer disks are optically thin, and the opacity is too small to 
account for the observed fluxes.
 
To compare with our models we show observations of Taurus stars taken from BSCG and BS91 
as stars in Figure \ref{fig_flux}. 
The observed spectral index is calculated as $n= \beta_p +2$,
where $\beta_p$, taken from \cite{BS91}, is the result of power law fits to the spectral energy
distributions between 1.3 mm and either 0.624~mm or 0.769~mm.  We ignore the effects of the different
wavelength limits on the spectral index; this has little effect on the comparison,
given the typical errors shown by the error bar in Figure \ref{fig_flux}  (BS91).
Figure \ref{fig_flux} shows that the disk models generally span the range of observed fluxes and 
spectral indexes larger  or similar to $n \sim 2$.

Figure \ref{fig_flux} also shows the flux and spectral index of 
flat irradiated disk models, with $a_{max}=$ 1 mm.   Here we have assumed that
the disks are geometrically flat, which corresponds to complete dust settling
to the disk midplane.  These disks are much colder at a given radius
than those in which dust and gas are well mixed up to several scale 
heights (flared disks), since they intercept a smaller fraction of the 
stellar radiation (e.g., \cite{H98}). 
The emission of these flat disks is dominated by optically thick and cold regions 
and both the flux and the spectral index are smaller ($n < 2$) 
than in the flared models.
Thus, the CTTS with $n < 2$ and a larger mm flux than the flat disk models 
could be intermediate cases, in which just a fraction of dust has been settled 
towards the midplane.

Panel (a) in figure \ref{fig_effects} shows the effect of changing the outer disk radius.
We have selected the parameters (i.e., $\Mdot$ and $\alpha$) of
the model with intermediate mass in Figure \ref{fig_flux}.
The models have $R_d=$ 50 AU (dashed line), $R_d=$ 100 AU (solid line) and
$R_d=$ 200 AU (dotted line), corresponding to masses $M_d \approx $ 0.02, 0.05 and 0.092 $\MSUN$, 
respectively (similar to the disk masses represented in Figure \ref{fig_flux}).
In general, the flux scales as $\sim R_d^{1/2}$,
as expected for an optically thin disk with $I_\nu \sim B_\nu \tau_\nu \propto R^{-3/2}$.
However,  for $a_{max} \sim$ 1 mm  and $R_d=50$ AU,
a larger fraction of the disk becomes optically thick which is reflected
by $n \rightarrow 2$.

Panel (b) in figure \ref{fig_effects} shows the effect of changing inclination angle.
For the same model with intermediate mass  and a radius $R_d=$ 100 AU ($M_d \sim 0.046 \ \MSUN$),
we change the inclination angle $\cos \ i =$ 0.25 (dashed), 0.5 
(solid) and 0.9 (dotted).
For $a_{max} < 150 \ \mu m$, the fluxes are similar
reflecting that the disk is mainly optically thin at $\lambda =1.3$ mm,
but the spectral index tends to decrease with the inclination of the disk,
because the relative contribution of the optically thick inner regions changes
at both wavelengths.
For $a_{max} >$ 1 mm, the flux roughly scales with $\cos \ i$, and the
spectral index is almost constant $n \sim 2.4$. 

Panel (c) in figure \ref{fig_effects} shows the flux vs spectral index of pole-on disk models 
with different mass accretion rates and masses, assuming no irradiation by the central star, i.e.
the disk heating is entirely due to viscous dissipation (Paper II). 
Each connected sequence corresponds to disk masses of $M_d=0.023, 0.046$ and $0.092 \ \MSUN$ 
and each point along a sequence corresponds to a different mass accretion rate 
(roughly differing by a factor 3).  We take $a_{max}=$ 1 mm and $p=3.5$ for all
models.  The flux increases with temperature (or $\Mdot$) and 
the spectral index is sensitive to both the disk temperature and mass. 
In these models the outer disk is very cold 
($T \sim 5- 15$ K)\footnote{We include the disk heating by cosmic rays and 
radioactive decay of $^{26}Al$, but the disk is so dense that it cools efficiently, 
and can reach temperatures lower than 10 K.} and the Rayleigh-Jeans limit is not valid. 
Unfortunately, since the optical depth is close to unity,  one cannot use 
neither the optically 
thin nor the optically thick limits of equation  (\ref{eq_jota}) to 
approximate 
the mm-wave fluxes emerging from the disk by simple expressions.
In addition, the disk temperatures are comparable to the cosmic 
background temperature 
and the optical depth of the disk is large, so the correction 
introduced when we substract   
$B_\nu(T_{bg})$ to the 
emergent intensity of the disk (c.f. \S \ref{sec_model}) 
becomes very important.
Besides this complications in describing analytically the results 
shown in Figure 
\ref{fig_effects},  the conclusions are simple.
Larger mass accretion rates for these pure viscous disk models result in larger mm-wave fluxes.
On the other hand, the smaller the disk mass, the smaller the flux and the
larger the spectral index, because the smaller the optical depth.
These models have $1 \lesssim n \lesssim 2.6$, but require
large mass accretion rates and small inclination angles to have
mm-wave fluxes consistent with the observations.

Panel (d) shows the same models as in panel (c), but including irradiation from the central star. 
All the models with different mass accretion rates at a given mass have almost the same flux 
at 1.3 mm and the same spectral index, because the stellar 
irradiation is the dominant heating source of the outer disk ($ R \gtrsim 20 \ \AU$). 

\section{Discussion}
\label{sec_disc}

We have shown that self-consistent disk
models with power-law distributions of grains, with $p \sim 2.5 - 3.5$,
and maximum sizes $a_{max} \sim 1$~mm, can reproduce many general features of
observed T Tauri disks, including SEDs, optical and near-infrared images, limits
on the frequency of disk-obscured systems, and mm-wavelength fluxes and spectral
indexes.  In contrast, our results of Paper II showed that ISM dust properties were
not in good agreement with observations.  
Our findings reinforce previous suggestions
(e.g., BS91, MN93) for grain growth in T Tauri disks with typical ages
of 1 Myr. We make these suggestions stronger,
however, since we aim to fulfill more
observational constraints.

The inference of substantial grain growth implies that there must be significant
uncertainties in deriving disk properties such as masses.  If dust agglomeration
has proceeded from typical interstellar sizes of $0.1 \ \mu$m to a 1 mm or larger,
there is no particular reason to assume that growth has proceeded 
to precisely the same value of $a_{max}$ in all systems.  While it may be possible
to argue that $a_{max}$ is at least $\sim 1$~mm on the basis of observed spectral
indexes (Figure 8), it is 
difficult 
to eliminate the possibility of growth
beyond this size from $n$ alone because $\beta$ approaches a constant for larger 
$a_{max}$ (Figure 3).  This uncertainty in dust size distributions (even assuming
$p$ can be constrained to lie between $\sim 2.5 - 3.5$) results in a corresponding
uncertainty in disk masses and other properties.

\begin{figure}
\plotone{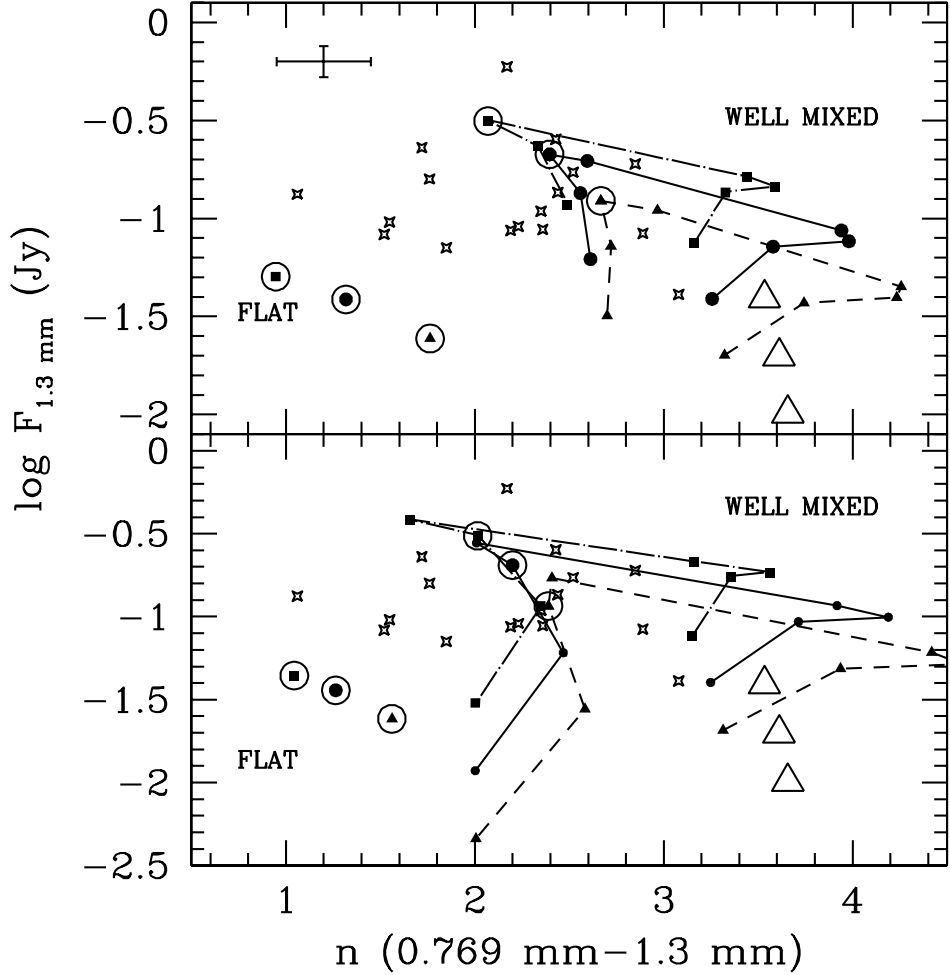}
\caption{Model disk flux at 1.3 mm vs. spectral index
$n$ between 0.769 mm and 1.3 mm (lines) compared with observed values (stars)
for Taurus from BSCG and BS91.
Lines connect models of the same mass:
$M_d \sim 0.023 \  \MSUN$ (dashed line),
$M_d \sim 0.046 \ \MSUN$ (solid line),
and $M_d \sim 0.092 \ \MSUN$ (dotted line).
Model values are calculated for maximum grain sizes
 $ a_{max}=$ 10, 100, 120, 150, 300 $\mu$m, 1 mm, 1cm and 10 cm,  
identified by dots along the lines.
For reference, the dot corresponding to
$a_{max} =$ 1 mm  is encircled.
Values for $a_{max} < 10 \ \mu$m  are similar to
those for  $ a_{max} = 10 \ \mu$m.
Upper panel: $p = 3.5$, lower panel: $p = 2.5$.
Typical 1-$\sigma$ errors are shown in the upper left corner
of the upper panel. 
The range of $\alpha$ covered by each sequence of models
is $\alpha = $ 0.02-0.04 ($M_d \sim 0.023 \  \MSUN$),
$\alpha = $ 0.01-0.02 ($M_d \sim 0.046 \ \MSUN$),
and $\alpha = $ 0.005-0.01 ($M_d \sim 0.092 \ \MSUN$). 
We also show the flux and spectral index for flat irradiated models 
with the same masses, and $a_{max}=$ 1 mm (circled dot, at the lower left 
corner) and for well mixed models with ISM-dust (big triangles). }
\label{fig_flux}
\end{figure}

\begin{figure}
\plotone{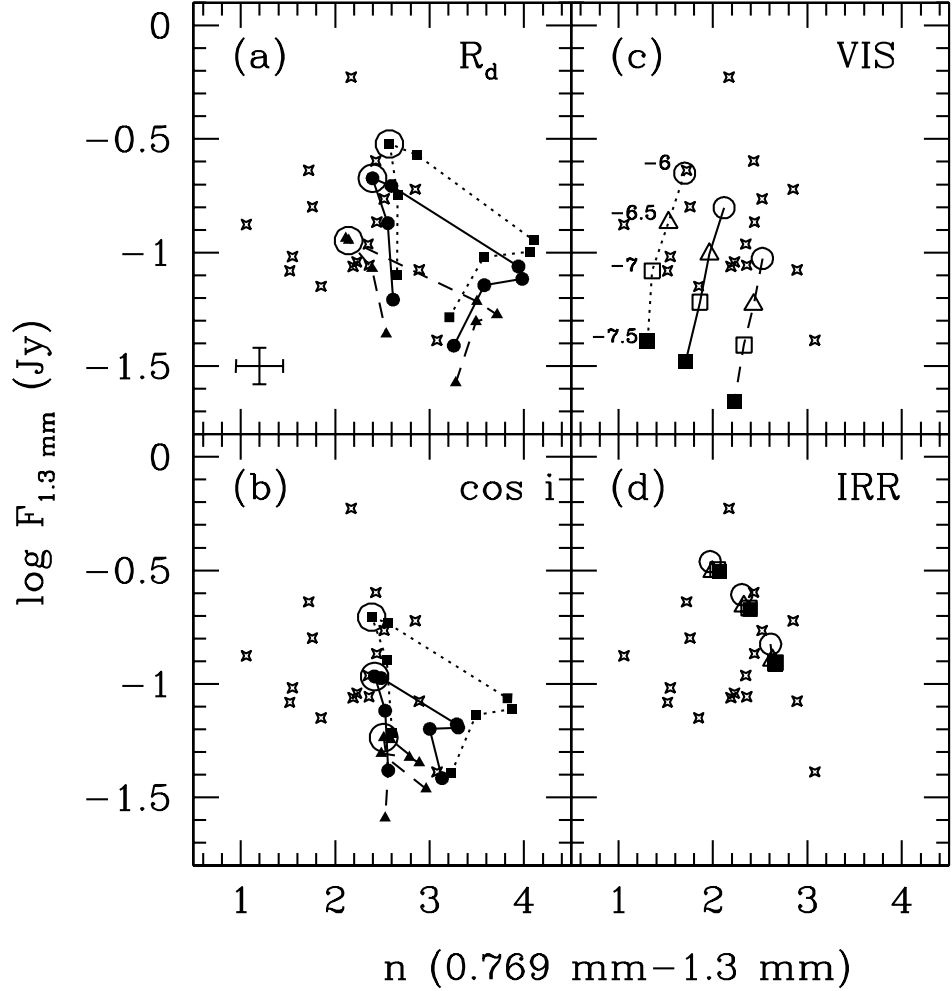}
\caption{Model disk flux at 1.3 mm vs. spectral index $n$ between 0.769 mm and 1.3 mm (lines)
for different disk radius, inclination angle and mass accretion rate.
(a) For the disk model with $p=$3.5, $M_d \sim 0.046 \ \MSUN$ and $R_d=100 \ \AU$, we change
the disk radius (and consequently the mass) keeping $\Mdot =3 \times 10^{-8} \ \MSUNYR$ and $\alpha=0.01$.
The plotted values are: $R_d=$50 AU (dashed line), 100 AU (solid line) and 200 AU (dotted line).
The values of $a_{max}$ and the observations are described in Figure \ref{fig_flux}.
(b) For the disk model with $p=$3.5, $M_d \sim 0.046 \ \MSUN$ and $R_d=100 \ \AU$, we
change the inclination angle, $\cos i=$ 0.25 (dashed line), 0.5 (solid line) and 0.9 (dotted line). 
(c) Non irradiated viscous disks. Each symbol corresponds to a different mass accretion rate 
(labeled by log $\Mdot$) and  the lines connect models with the same mass, 
$M_d=0.092$ (dotted line), $0.046$ (solid line) and $0.023 \ \MSUN$ (dashed line). 
For all these models $a_{max}=1 \ mm$, $p=3.5$ and $i=0^\circ$.
(d) The same models  shown in panel (c), irradiated by the central star (assuming dust and gas well mixed).}
\label{fig_effects}
\end{figure}

Limiting dust opacity variations by constraining $a_{max}$ to lie between $100 \  \mu$m 
and 1 cm, our disk models suggest that disk masses between $\sim 10^{-2} \MSUN$
and $\sim 10^{-1} \MSUN$ are consistent with the data.  The upper end of this mass range
encompasses disks subject to gravitational instability.
Specifically, the Toomre parameter $Q$ (e.g., D'Alessio et al. 1999b) at $R_d = 100$~AU
is $Q \sim 1$ for the $M_d = 0.05 \MSUN$ models and $Q \sim 0.5$ in the $M_d = 0.1 \MSUN$ models,
suggesting that the former model is close to instability and the latter is strongly unstable.
Thus, the possibility of gravitational instabilities in at least some T Tauri disks 
cannot be dismissed.  
In this regard,
models shown in panels (c) and (d) of Figure \ref{fig_effects} 
indicate that
observed high mm fluxes and low spectral indices of some of the
stars could be explained by high mass disks with high mass accretion
rates, as long as their outer disks were not heated by irradiation.
Interestingly, some of the stars with $n < 2$ 
(CW Tau, DR Tau and DL Tau)  are high mass accretion rate
stars with $\Mdot \sim 2 \times 10^{-7}-8 \times 10^{-6} \ \MSUNYR$ (\cite{HEG}). 
We may speculate that one
possible interpretation is that these disks may be gravitationally
unstable and the outer disks are shadowed from the
star by density perturbations produced by
the development of the instability, as models suggest 
(e.g., Pickett et al. 2000, Nelson et al. 2000). 
If one assumes that the disk masses are always smaller than the gravitational stability limit, then
the high mm fluxes shown in Figure \ref{fig_flux} cannot be understood if
grains have grown beyond about 1 cm in size.  However, growth to larger sizes 
could be accommodated in the inner disk; 
these are optically thick regions and the resulting emission is insensitive to their total mass.

The models predict that less-luminous disks are more likely to be 
optically thin and
thus exhibit spectral indexes which reflect the intrinsic wavelength 
dependence
of the dust opacity much more closely.  There are many nearby T Tauri 
stars whose 1.3 mm fluxes 
lie in the range of $\sim 15-30$~mJy (\cite{OB95}) for which
mm-wave spectral indexes have not been published.  Observations of
spectral indexes for these fainter objects could more clearly demonstrate 
particle growth.
However, because the spectral index tends to a limiting value when 
$a_{max}$ becomes large,
it will be difficult to tell
whether low mm-wave emission is due to low intrinsic disk 
mass or particle growth, 
unless mid- and far-IR fluxes and scattered light images
are used to disentangle disk mass and grain sizes.

In principle, geometrically-thinner disks could be the result of 
dust settling alone.
However, our analysis of the mm-wave fluxes and spectral 
indexes indicates that grain growth
is essential.  Because the growth of particles reduces the opacity at 
short wavelengths,
it is difficult to distinguish between the effects of settling and 
grain growth on observed
SEDs and images.  The case of HK Tau/c illustrates how one may arrive 
at very different
conclusions about disk masses depending upon the dust properties assumed.

Of course, grain growth should be accompanied by settling; larger 
particles tend
to fall more rapidly toward the disk midplane 
(e.g.,\cite{WC93}, \cite{W97}).  \cite{MN95} 
argue that settling
has to occur in some T Tauri disks on the basis of their infrared SEDs.  Two of
our results also support the idea of settling.  First, our models 
generally cannot
explain the T Tauri stars with moderately low mm-wave fluxes and small values of $n$
(e.g., Figure \ref{fig_flux}).
  Flat disks produce lower mm-wave fluxes
and smaller spectral indexes, because the reduced irradiation fluxes reduce the outer
disk temperatures.  
Our detailed calculations show that  
geometrically-flat disks result in too small values of flux and $n$, but the results
shown in Figure Figure \ref{fig_flux} suggest that disks with intermediate thicknesses (i.e., some settling
of dust) could explain the T Tauri disks found in the lower-left hand region of
the flux-spectral index diagram.

Second, our models with $a_{max} \sim 1$~mm and $p = 2.5 - 3.5$ 
have no silicate feature,
in contrast with observations of many T Tauri stars 
(\cite{CW85}, \cite{NMB}).
In the context of our disk models, in which the relative grain
size distribution is independent of height and radius, putting
more of the total dust mass into large grains (grain growth)
reduces the geometrical disk thickness at large radii,
in better agreement with observations, but results in too few
small grains at 1 AU ($T \sim$  100-300 K) to produce 10 $\mu$ 
silicate emission.
We could improve the agreement with observations
by having more small grains at 1 AU than at large radii.
If the grain size distribution is independent of height, this
would imply that there has been less grain growth at small radii
than at large disk radii.  It seems much more likely that there
has been faster grain growth at small radii, but that the
large grains have also settled more toward the midplane in the inner
disk.  This settling could result in a much larger fraction of
small grains in the upper atmospheric layers at ~ 1 AU, which
is the region where the silicate emission is formed (\cite{CMPD}; \cite{CG97}).

It is therefore clear that improved disk models for T Tauri 
stars will have to incorporate
both grain growth and settling in a physically self-consistent manner,
 accounting for the 
radial and vertical dependences of both process.   We will develop
such models in a forthcoming paper.  

This work was supported in part by the Visiting 
Scientist Program of the Smithsonian
Institution, by NASA grant NAG5-9670, by Academia Mexicana de Ciencia and
Fundaci\'on M\'exico-EEUU para la Ciencia, by CONACyT grant J27748E, M\'exico.
We thank Luis Felipe Rodr\'\i guez for important advice 
about observational procedures in the mm-wave range, Karl Stapelfeld 
for sending us the image of HK Tau/c, Mordecai Mark Mac Low and Javier Ballesteros for useful discussions during this work.
We also thanks Robbins Bell for a prompt and useful referee report.
PD acknowledges the hospitality of the Astrophysics Department 
of Columbia University and a fellowship of DGAPA-UNAM, M\'exico.

\newpage

\begin{figure}
\plotone{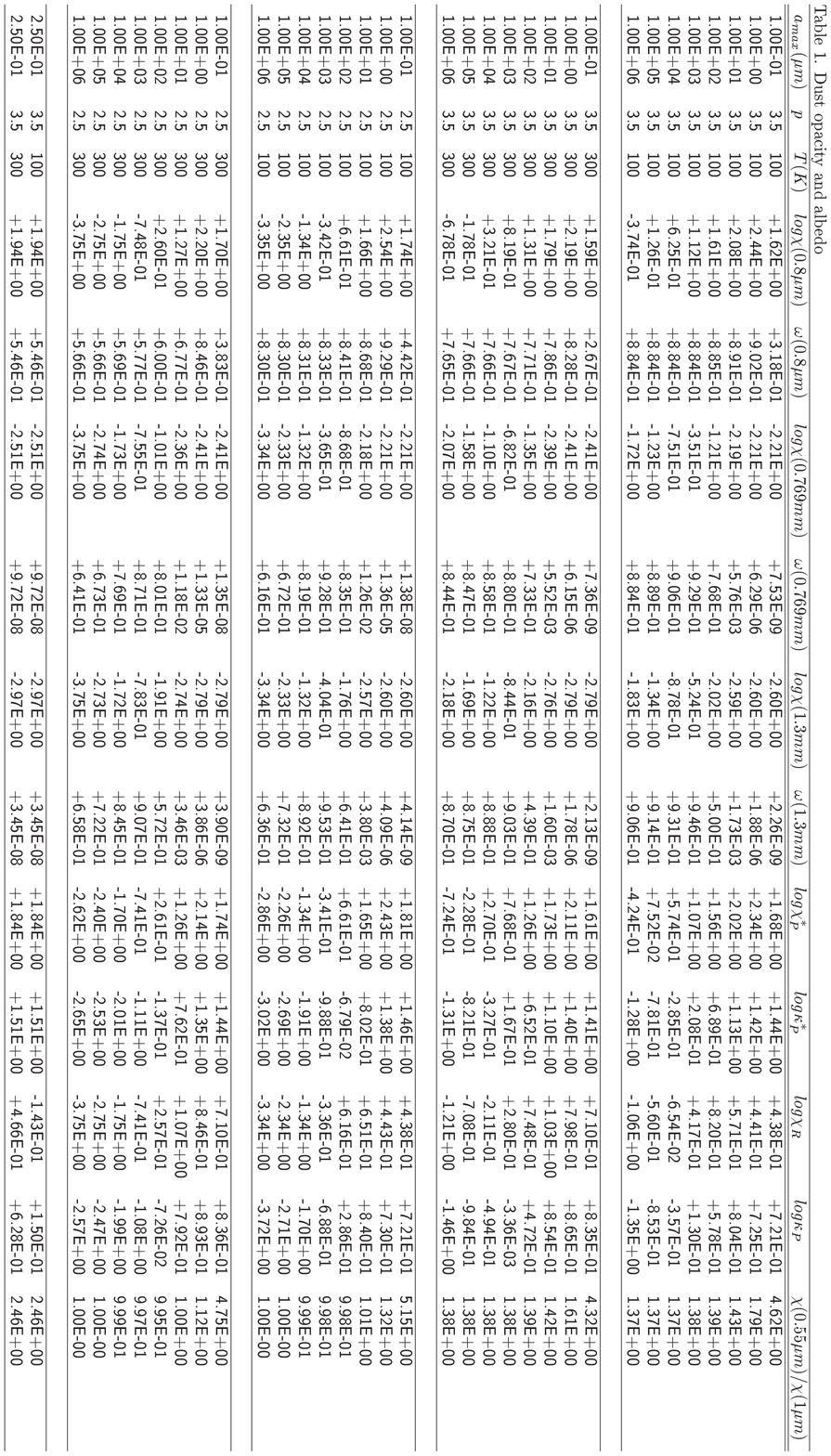}
\end{figure}

\end{document}